\documentclass[twocolumn,preprintnumbers,amsmath,amssymb,superscriptaddress]{revtex4}
\usepackage{graphicx}% Include figure files
\usepackage{dcolumn}% Align table columns on decimal point
\usepackage{bm}% bold math

\def\ltsima{$\; \buildrel < \over \sim \;$}
\def\simlt{\lower.5ex\hbox{\ltsima}}
\def\gtsima{$\; \buildrel > \over \sim \;$}
\def\simgt{\lower.5ex\hbox{\gtsima}}

\let\sec=\section

\def\[{\begin{equation}}
\def\]{\end{equation}}
\def\m@th{\mathsurround=0pt }
\def\eqalign#1{\null\,\vcenter{\openup1\jot \m@th
 \ialign{\strut\hfil$\displaystyle{##}$&$\displaystyle{{}##}$\hfil
 \crcr#1\crcr}}\,}

\begin{document}
\title{On lensing by a cosmological constant}

\author{Fergus Simpson}
 \email{frgs@roe.ac.uk}
\author{John A. Peacock}
\author{Alan F. Heavens}
\affiliation{SUPA, Institute for Astronomy, University of
Edinburgh, Royal Observatory, Blackford Hill, Edinburgh EH9 3HJ}

\date{\today}
\newcommand{\ud}{\mathrm{d}}

\begin{abstract}
Several recent papers have suggested that the cosmological
constant $\Lambda$ directly influences the gravitational
deflection of light. We place this problem in a cosmological
context, deriving an expression for the linear potentials which
control the cosmological bending of light, finding that it has no
explicit dependence on the cosmological constant. To explore the
physical origins of the apparent $\Lambda$-dependent potential
that appears in the static Kottler metric, we highlight the two classical effects which lead to the aberration of light. The first relates to the
observer's motion relative to the source, and encapsulates the
familiar concept of angular-diameter distance. The second term,
which has proved to be the source of debate, arises from cosmic
acceleration, but is rarely considered since it vanishes for
photons with radial motion. This apparent form of light-bending
gives the appearance of curved geodesics even within a flat and
homogeneous universe. However this cannot be construed as a real
lensing effect, since its value depends on the observer's frame of
reference. Our conclusion is thus that standard results for
gravitational lensing in a universe containing $\Lambda$ do not
require modification, with any influence of $\Lambda$ being
restricted to negligible high-order terms.

\end{abstract}
\maketitle

\sec{Introduction} Conventional wisdom (e.g.
\cite{1983PhLA...97..239I,2006astro.ph..1044F}) states that the
cosmological constant plays no direct role in gravitational
lensing, other than the inevitable modification to the angular
diameter distance. This is reinforced by the intuition that
lensing is sourced by inhomogeneities in the density field,
whereas the cosmological constant is wholly uniform.

This position was challenged by Rindler \& Ishak
\cite{2007PhRvD..76d3006R}, who presented a term associating the
cosmological constant with a diminished bending angle for a
photon. This was followed by a further two papers
\cite{2007arXiv0710.4726I,2008arXiv0801.3514I}  analysing this
phenomenon in greater detail. Indeed the former claims to place
observational constraints on the value of $\Lambda$ based on
applying this result to strong lensing by clusters, in a
`Swiss-cheese' model, where the matter in a spherical vacuole
collapses to the centre to form the lensing object. Although the
effects are relatively small, they are certainly large enough to
be important in next-generation applications of lensing as a tool
for precision cosmology. However, opinion seems divided as to
whether the Ishak-Rindler analysis is correct: Park
\cite{2008arXiv0804.4331P} and Khriplovich \& Pomeransky \cite
{2008arXiv0801.1764K} have shed doubt on these calculations,
although Sch\"{u}cker
\cite{2008A&A...484..103S,2008GReGr.tmp...86S,2008arXiv0805.1630S},
and Lake \cite {2007arXiv0711.0673L} are in agreement. Work by
Gibbons et al. \cite{GibbonsWerner} explore the properties of the
Kottler optical metric, while Sereno \cite
{2008PhRvD..77d3004S,2008arXiv0807.5123S} revealed a different
term contributing to the deflection angle.

In this work we aim to clarify the source of these discrepancies
and to investigate the bending of light in an expanding Universe.
%our conclusion is that there is no evidence to suggest that
%$\Lambda$ plays any direct role in the deflection of photons.
In \S\ref{sec:vac}, we translate the metric inside a vacuole from
the static Kottler \cite{1918AnP...361..401K} form to a perturbed
Friedmann-Robertson-Walker (FRW) metric. We do not exclude a
contribution of $\Lambda$ to the lensing equations at some level,
but show that the linear potential is unaffected by $\Lambda$,
with the apparent $\Lambda r^2/3$ contribution appearing as a
consequence of the choice of a static metric. We verify this with numerical solutions in \S\ref{sec:numerics}.

The remainder of this work aims to clarify the physical
interpretation of the apparent light bending. We revisit the
analysis of Ishak \cite{2008arXiv0801.3514I} in
\S\ref{sec:angles}, before extending this to evaluate the photon's
deflection angle from different perspectives within the Kottler
metric. The source of the extra term is revealed in
\S\ref{sec:origins}, and its relation to the angular-diameter
distance is outlined in \S\ref{sec:worked}. Final discussions are
presented in \S\ref{sec:conc}.

\sec{Vacuole model in the Newtonian gauge} \label{sec:vac}

We now consider the Ishak--Rindler vacuole from the point of view
of the standard approach to cosmological perturbations, as
described by e.g. Dodelson \cite{2003moco.book.....D} or Mukhanov
\cite{2005pfc..book.....M}. Our goal is to find an explicit linear
expression for the perturbing potentials responsible for the
cosmological bending of light within the vacuole model. In order
to avoid coordinate-dependent artefacts, one looks for
gauge-independent measures of inhomogeneity; in practice, this is
achieved by working in the Newtonian gauge. Scalar metric
fluctuations are then described by scalar potentials, $\Phi$ and
$\Psi$, which act to modify the Robertson--Walker metric:
\[
ds^2 = (1+2\Phi)\, dt^2 - a^2(t)\, (1-2\Psi)\, (d\chi^2 + \chi^2
d\psi^2).\label{eq:FRW}
\]
We take $c=G=1$ throughout. $\chi$ is comoving radius, and $d\psi$
is an element of angle on the sky. We also restrict attention to
the case of a flat universe, and no anisotropic stresses, so that
$\Psi=\Phi$. We will always be interested in the case where the
fluctuations causing lensing are well within the horizon, in which
case the potential $\Phi$ obeys the Poisson equation, sourced by
the fractional matter fluctuation $\delta_m$. In this apparatus, a
homogeneous density from $\Lambda$ appears only implicitly,
through its contribution to the scale factor $a(t)$.
Conventionally, light deflection would be computed by integrating
twice the component of $-{\bf\nabla}\Phi$ perpendicular to the
line of sight, and the conclusion would be that $\Lambda$ has no
direct lensing effect. Clearly this is true in a homogeneous
universe that contains $\Lambda$, since the FRW metric defines the
path of unperturbed light rays. Indeed, no true lensing can arise
from a homogeneous background: the photon would require a
preferential direction in which to bend -- and doing so would
break the symmetry of the cosmology.

How does the perturbed FRW metric compare with the exact Kottler
metric inside the vacuole? The comparison can only be made if we
understand the relation between the coordinates used in the two
forms. The key to doing this is the transverse part of the metric,
which would be $-r^2 d\psi^2$ in the Kottler form:
\[
ds^2 = f(r)\, dT^2 - f(r)^{-1} dr^2  - r^2 d\psi^2,
\]
for some time coordinate $T$, and where
\[
f(r) = 1 -{2m\over r} -{\Lambda r^2\over 3}.\label{eq:Kottler}
\]
This means that the proper radius in the perturbed FRW form is, to
first order in $\Phi$,
\[
r \equiv (1-\Phi) a(t) \chi, \label{eq:perturbed}
\]
and the metric is
\[
ds^2 = (1+2\Phi)\, dt^2 - a^2(t)\, (1-2\Phi)\, d\chi^2  - r^2
d\psi^2.
\]
In order to eliminate $d\chi$, we must differentiate the
definition of $r$. To first order in $\Phi$, this gives
\[
a^2(1-2\Phi)d\chi^2 = \left[ (1+\chi \Phi')\, dr - Y\,
dt\right]^2,
\]
where $Y\equiv \dot a\chi(1-\Phi) - a\chi\dot\Phi + \dot a \chi^2
\Phi'$, where we have defined $\Phi' \equiv \ud \Phi / \ud \chi$ and $\dot\Phi \equiv \ud \Phi / \ud t$. Note that we always work to first order in the
perturbation, so that e.g. $(1-\chi \Phi')^{-1}$ can be replaced
by $(1+\chi \Phi')$. Note also that differentiating the definition
of $r$ introduces $\dot a$: $\Lambda$ enters at this point, since
it is related to  $\dot a$ via the Friedmann equation.

In order to eliminate the $dr\, dt$ cross term, we define the
time coordinate $dT = A\, dt + B\, dr$, and solve for $A$ and $B$
by requiring that the metric be written in the desired Kottler
form,
\[
ds^2 = f(r)\, dT^2 - f(r)^{-1} dr^2  - r^2 d\psi^2.
\]
Solving this problem as written gives
\[
f = { 1+2\Phi - Y^2 \over (1+\chi \Phi')^2(1+2\Phi) }.
\]
To first order in $\Phi$, our expression for $f$ is
\[
\eqalign{ f = 1 &-2\chi\Phi' - \dot a^2\chi^2 (1-4\Phi -
2\chi\Phi') \cr &+2 \dot a a\chi^2\dot\Phi -2 \dot a^2 \chi^3
\Phi' \cr
%+2 \dot a a\chi^2\dot\Phi -2 \dot a^2 \chi^3 \Phi' \nonumber\\
% &= 1-2\chi\Phi' - \dot a^2\chi^2 (1-4\Phi)
%+2 \dot a a\chi^2\dot\Phi \cr
 = 1 &-2\chi\Phi' - \dot a^2\chi^2 \left[1-2\Phi
\left(\frac{\partial \ln |\Phi|}{\partial \ln a}+2\right)\right].
}\label{eq:main}
\]
We proceed initially by ignoring the terms proportional to $\Phi$
in the square brackets, in comparison to unity. This is justified
because the vacuole should be small compared with the Hubble
length: $\dot a \chi \ll 1$. Thus, $\dot a^2\chi^2\Phi$ will be
negligible in comparison with $\chi \Phi' = \Phi (\partial\ln
\Phi/\partial\ln \chi)$. We show below that this assumption yields
a consistent solution for $\Phi$, with the neglected term shown to
be second order.

In practice, therefore, the perturbed FRW metric reduces to the
Kottler form with
\[
f = 1-2\chi\Phi' - \dot a^2\chi^2.
\]
The Kottler metric is expressed in terms of $r=(1-\Phi)a(t)\chi$,
but we have already treated $\dot a^2\chi^2\Phi$ as negligible in
deriving this expression for $f$, so the same level of
approximation allows us to set $r=a(t)\chi$ here:
\[
f = 1-2r\Phi'/a - (\dot a^2/a^2)r^2.
\]
The Friedmann equation says that
\[
\dot a^2/a^2 = \Lambda/3 + 2m/R^3,
\]
where $R=a(t)R_v$ is the proper radius of the vacuole of fixed
comoving radius $R_v$. Using this and the Kottler metric yields
\[
f = 1-2r\Phi'/a -2mr^2/R^3 - \Lambda r^2/3.
\]
Note that the Friedmann equation has yielded a term $- \Lambda
r^2/3$, which will cancel the corresponding term in the Kottler
expression for $f(r)$.

Recalling that $\Phi'$ denotes the derivative of $\Phi$ with
respect to comoving radius, we solve this using the Kottler form
for $f(r)$, equation (\ref{eq:Kottler}), which requires
\[
\Phi' = a\left({m\over r^2} - {mr\over R^3}\right) = {m\over
a\chi^2} - {m\chi\over aR_v^3},
\]
where again the error in writing $r=a(t)\chi$ is of second order
in $\Phi$. The solution is
\[
\Phi =-{m\over a\chi} - {m\chi^2\over 2 a R_v^3} + {3m\over 2 a
R_v} = -{m\over r} - {mr^2\over 2R^3} + {3m\over 2R} ,
\label{solution}
\]
where the additive constant is determined by requiring $\Phi=0$ at
the boundary of the vacuole. This expression for $\Phi$ agrees
with what one would expect from a simple Newtonian calculation
with a point mass and a spherical vacuole underdensity. We can
dispose of the technical issue that for a point mass, $\Phi
\rightarrow -\infty$ as $\chi\rightarrow 0$, by considering a
spherical mass of finite radius, and appealing to Birkhoff's
theorem so that our solution for $\Phi$ applies in the vacuole
outside the mass.

To verify that we have a consistent linear solution for $\Phi$, we
note that $\partial\ln|\Phi|/\partial \ln a=-1$, and substitution
into (\ref{eq:main}), retaining the full relation
(\ref{eq:perturbed}) to linear order in $\Phi$, gives a complete
cancellation of the $\Lambda$ terms:
\[
\Phi' = \frac{m}{a\chi^2} - {m\chi\over a R_v^3} +
\Phi\left({m\over a \chi^2} + {2m \chi\over a R_v^3}\right).
\]
The last term on the right, which we neglected, is indeed seen to
be second-order in $\Phi$.

Having now described the vacuole metric as a perturbed FRW metric,
we find no evidence for $\Lambda$-dependence in the linear
peculiar gravitational potential, and the standard cosmological
lensing results follow. The potential-like term $\Lambda r^2/3$ in
the Kottler $f(r)$ arises simply by virtue of the introduction of
$\dot a$ in the coordinate transformation from the FRW form, plus
the fact that $\dot a$ is related to $\Lambda$ through the
Friedmann equation. But this term does not arise from the true
potential $\Phi$, and thus it should not be taken to cause
lensing. From this point of view, it seem fair to assert that the
appearance of a lensing effect from $\Lambda r^2$ in the
Ishak--Rindler analysis is purely a gauge artefact.

Our expression for the potential $\Phi$ is correct only to lowest
order, and we have neglected corrections of order $(H^2r^2)\Phi$.
Nevertheless, it is clear that the disputed $\Lambda r^2/3$ term
is of an altogether larger magnitude. In most of the volume of the
vacuole, $r\sim R$ and $\Phi \sim m/R$. The ratio between the
disputed term and $\Phi$ is then $\sim \Lambda R^3/m$, which is of
order the ratio between the vacuum and matter densities -- i.e. an
order unity correction at the present epoch. While $\Lambda$ may
appear in higher order corrections to $\Phi$, it is clear that
such corrections cannot involve a $\Lambda r^2/3$ term in the
potential.

\sec{Numerical Solutions} \label{sec:numerics}

We now derive expressions for the linear peculiar gravitational potential for a spherical vacuole in a universe containing arbitrary densities of non-relativistic matter and a cosmological constant, in which the mass $m$ in the vacuole is concentrated at the centre.  Otherwise, the universe is assumed to be homogeneous and isotropic.

The perturbed FRW metric is written in the Newtonian gauge as
\begin{equation}
ds^2 = (1+2\Phi) dt^2 - a^2(t)(1-2\Phi)[d\chi^2 + S_k^2(\chi)d\psi^2]
\end{equation}
where $S_k(\chi)=R_0 \sin(\chi/R_0), \chi, R_0\sinh(\chi/R_0)$ for $k=1,0,-1$, and $R_0 = (1/H_0)\sqrt{k/(\Omega-1)}$.

We wish to find $\Phi(\chi,a)$ by coordinate transformation from the known Kottler metric (\ref{eq:Kottler}) in terms of static coordinates $r$ and time coordinate $T$. Equating the coefficients of $d\psi^2$ gives (working always only to linear order in $\Phi$)

\begin{equation}
r(\chi,a) = a(t)(1-\Phi)S_k(\chi).
\end{equation}

\noindent We also have $T=T(\chi,a)$ and hence $dT = T_t dt + T_\chi d\chi$ where $T_t = \partial T/\partial t$ etc.  Substitution into the Kottler metric (\ref{eq:Kottler}), equating coefficients of $dt^2$, $d\chi^2$ and $dt d\chi$, eliminating $T_t$ and $T_\chi$ gives
\begin{equation}
r_\chi^2  (1+2\Phi) - a^2(1-2\Phi)r_t^2 - a^2  f(r) = 0.
\end{equation}
Since
\begin{eqnarray}
r_\chi &=& a(1-\Phi)C_k - a \Phi_\chi S_k\nonumber\\
r_t &=& \dot a (1-\Phi)S_k - a \Phi_t S_k,
\end{eqnarray}
where $C_k(\chi) \equiv \cos(\chi/R_0), 1, \cosh(\chi/R_0)$ for $k=1,0,-1$, we obtain (to $O(\Phi)$),

\begin{widetext}
\[
1 - \frac{2m}{a S_k }(1+\Phi) - \frac{a^2\Lambda S_k^2}{3}(1-2\Phi)  = C_k^2 - 2 \Phi_\chi S_k C_k - \dot a^2 S_k^2 (1-4\Phi) + 2\dot a^2 a S_k^2 \Phi_a.
\]

\noindent Since $1-C_k^2 = k S_\chi^2 /R_0^2$, we have the equation for $\Phi(\chi,a)$:

\[
\eqalign{
\Phi_\chi = \frac{S_k}{C_k}\left[\frac{-k}{2 R_0^2}  - \frac{\dot a^2}{2}(1-4\Phi-2a \Phi_a)  + \frac{m}{a S_k^3}(1+\Phi) + \frac{\Lambda a^2}{6}(1-2\Phi)\right].}
\]
Substituting Friedmann's equation $\dot a^2/a^2 = 8\pi \rho_m/3 + \Lambda /3-k/(a^2 R_0^2)$ gives

\[\Phi_\chi = \frac{S_k}{C_k}\left[- \frac{4\pi m}{3 V_b a}(1-4\Phi-2a \Phi_a) + \frac{m}{a S_k^3}(1+\Phi)   + \left(\frac{\Lambda a^2}{3} - \frac{k}{R_0^2}\right)(\Phi+a\Phi_a)\right],
\]
\end{widetext}

\begin{figure}
\centering
\includegraphics[width=80mm]{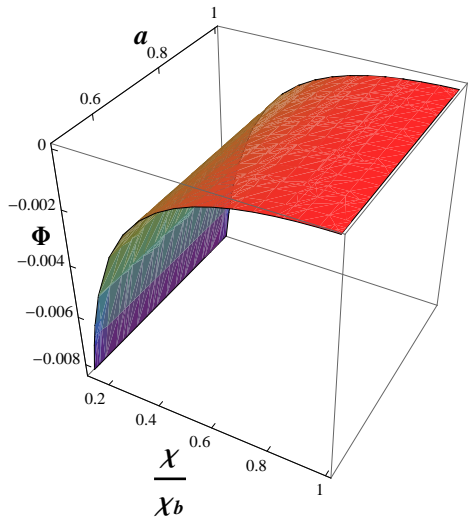}
\caption{The potential $a\Phi(\chi/\chi_b,a)$ of an open universe, with $\Omega_m=0.27, \Omega_\Lambda=0$. Here we have taken the vacuole size to be $\chi_b=0.1$.}
\label{fig:matter}     
\end{figure}

\begin{figure}
\centering
\includegraphics[width=80mm]{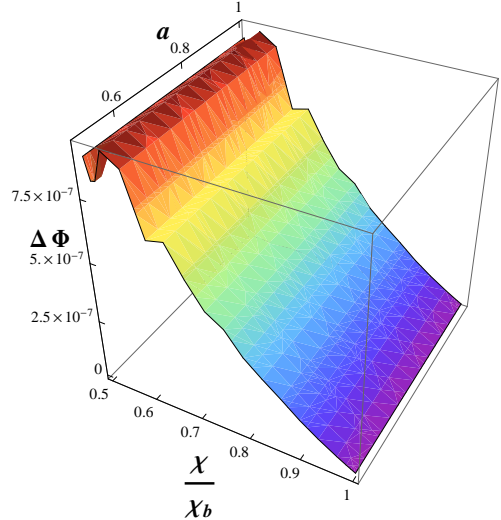}
\caption{The difference between the matter-only potential presented in Figure \ref{fig:matter}, and that of a cosmology with both matter and a cosmological constant:  $\Omega_m=0.27$,  $\Omega_\Lambda=0.27$ .}
\label{fig:difference}     
\end{figure}

\noindent where the comoving volume of the vacuole in terms of its boundary coordinate $\chi_b$ is given by

\[
V_b = \pi R_0^3 \left[ \sinh \left( \frac{2 \chi_b}{R_0} \right) - \frac{2 \chi_b}{R_0} \right] ,
\]

\noindent resulting in a vacuole of mass 

\[
m =  \frac{3 V_b \Omega_m h^2}{8 \pi}  .
\] 

\noindent Note that when modifying the cosmological parameters such that $R_0$ changes, we iteratively adjust the vacuole radius $x_b$ in order to maintain a constant enclosed mass $m$.

Figure \ref{fig:matter} shows Mathematica solutions for $a\Phi$ against $\chi/\chi_b$ (plotted from 0.1 to 1) and $a$ (plotted from 0.5 to 1) for $\Omega_m=0.27$ and $\Omega_\Lambda=0$, with $\chi_b = 0.1 c H^{-1}_0$. The leading-order term is given by

\[
\Phi(r) \simeq  - \frac{m}{r}  .
\]

\noindent When introducing a contribution from the cosmological constant, Ishak \& Rindler predict a change in the potential given by a new term

\[
\eqalign{
%\Phi(r) & =  - \frac{m}{r} - \frac{\Lambda r^2}{12} \cr
%\Delta \Phi(r) &= \Omega_\Lambda  h^2 \frac{(r-r_b)^2}{4}
\Phi_{\Lambda}(r) & = - \frac{\Lambda r^2}{12} \cr
 &= - \Omega_\Lambda  h^2 \frac{(r-r_b)^2}{4}  ,
}
\]

\noindent where the $r_b$ term arises due to the boundary condition imposed at the edge of the vacuole.  For a cosmology with $\Omega_\Lambda=0.27$ this leads us to predict $\Phi_\Lambda(r_b/2) =  -1.7 \times 10^{-4}$, where we have adopted  $r=r_b/2$ as the benchmark value.  However the signal in Figure \ref{fig:difference} is more than two orders of magnitude smaller than this, an amplitude consistent with the $O(\Phi^2)$ terms we discarded earlier. 

In the following sections we will explore why the Ishak-Rindler term does not manifest itself in the potential.

\sec{Deflection in the static metric} \label{sec:angles}

Now let us see how the above section can be made consistent with
the Ishak and Rindler computation of deflection within a static
Kottler metric.

%If we are to assess the deflection angle of a photon, first we
%must carefully define the measurement process for an angle. The
%coordinate angle relative to the radial direction $\hat{r}$ is
%simply
%
%\[ \label{eq:coorangle}
%\tan \theta_c = r \frac{\ud \phi}{\ud r},
%\]
%
%\noindent which is related to the physical angle $\theta_p$ by the
%radial and angular components of the metric
%
%\[ \label{eq:trueangle}
%\tan \theta_p = \sqrt{\frac{g_{\phi \phi}}{g_{rr}}} \frac{\ud
%\phi}{\ud r} .
%\]

We begin by reassessing the treatment outlined by Ishak
\cite{2008arXiv0801.3514I}, which relies on superposing two
metrics, and applying classical Newtonian dynamics, a valid
approximation provided we restrict ourselves to the weak field
regime. For a flat universe consisting of non-relativistic matter
and a cosmological constant, its evolution may be well described
by an appropriate choice of potential. This was applied by Ishak
\cite{2008arXiv0801.3514I} to evaluate the deflection angle of a
photon by considering the gradient of the Newtonian potential.
\[ \label{eq:delf}
\alpha  =  \int \nabla_\bot (\Phi + \Psi) \, \ud x  ,
\]
\noindent where we integrate along the path of the photon. The
potentials $\Phi$ and $\Psi$ are extracted from the space and time
components of the metric, and differ from the FRW potentials in equation (\ref{eq:FRW}).

Ishak superposes the Schwarzschild and de-Sitter metrics in static
coordinates,
\[
d s^2 =  \left(1+2\Phi\right) d t^2 -
\left( 1-2\Psi\right) \left( d r^2 + r^2 d \Omega^2\right)
\]
%\noindent where $\psi = \Lambda r^2/12$,
leading to the potentials (see eg
\cite{2003CQGra..20.2727K,2008arXiv0801.3514I}).
\[ \label{eq:potentials}
\eqalign{ \Phi \simeq - \frac{m}{r} - \frac{\Lambda r^2}{6} , \cr
\Psi \simeq - \frac{m}{r} + \frac{\Lambda r^2}{12}.}
\]
The mass terms are only an
approximation, but are accurate to first order provided $m/R \ll
1$. It is the terms involving the cosmological constant to which
we will pay the greater attention.

In the frame of the lens, the gradients of the potentials are

\[ \label{eq:alpha1}
\eqalign{ \nabla_\bot \Phi  &= \frac{mR}{r^3} - \frac{\Lambda
R}{3} ,\cr \nabla_\bot \Psi  &= \frac{mR}{r^3} + \frac{\Lambda
R}{6} . }
\]
\noindent Therefore a photon passing within a distance $R$ becomes
deflected by

\[
\eqalign{\alpha & =  \int_{-x_b}^{x_b} \nabla_\bot (\Phi + \Psi)
\, \ud x \cr & =  \left(\frac{4M}{R}-\frac{\Lambda R r_b}{3}
\right) \sqrt{1 - R^2/r_b^2}} \label{eq:alpha2}
\]

\noindent where $r=\sqrt{R^2+x^2}$, and $r_b$ is the radius of the
vacuole. This result corrects for the discontinuity which arises
at the boundary $R=r_b$ from Ishak \cite{2008arXiv0801.3514I}.

\begin{figure}[t]
\includegraphics[width=80mm]{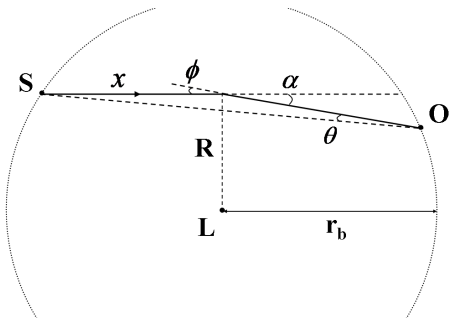}
\caption{The weak gravitational lensing of light by a vacuole,
where all matter within a sphere of radius  of radius $r_b$ is
collected to a central mass. The observer is located at $O$.}
\label{fig:fig1}
\end{figure}

From (\ref{eq:alpha2}), the first term within the parentheses is
readily recognisable as the conventional result for gravitational
lensing.
%As an aside, we note that this approach provides a good
%illustration of where the extra factor of two arises when
%switching from Newtonian physics to General Relativity. A
%non-relativistic object travelling along a timelike geodesic is
%only deflected by the time-time component of the metric, $\Phi$.
%But since the photon traverses a null geodesic it is equally
%affected by the space-space component, $\Psi$, thus the magnitude
%of the effect is doubled.
%
However it is the second term which is of greater interest. Why
does the cosmological constant now appear to reduce the value of
$\alpha$? Part of the reason is that, unlike the mass $m$, the
cosmological constant has a potential which appears centred on
whichever frame we choose. If the potential associated with the
cosmological constant is now centred on the observer at $O$, then
for the limit of a very weak lens the photon's motion is purely
radial, with no component perpendicular to the potential's
gradient, and thus the integral in (\ref{eq:alpha2}) trivially
vanishes. Yet we have not fully resolved the anomaly, since for
non-negligible deflection angles, the photon's path \emph{does}
have a transverse component before reaching the lens, as
illustrated in Figure \ref{fig:fig1}. Note however that the
observable, $\theta$, remains constant.

If the conventional deflection angle induced by the matter in the lens, $\alpha_m$, is small, then our modified deflection angle requires the inclusion of an extra term given by

\[
\eqalign{ \alpha_\Lambda & =   \int_{0}^{r_S} \nabla_\bot (\Phi +
\Psi)\, \ud x \cr & =  \int_{0}^{r_S} \nabla_\bot (-\frac{\Lambda
r^2}{6} + \frac{\Lambda r^2}{12})\, \ud x \cr & = \int_{r_L}^{r_S}
-\frac{\Lambda}{6} r  \sin \phi \,\ud x \cr & = \int_{r_L}^{r_S}
-\frac{\Lambda}{6} r_L \sin \alpha_m\, \ud x \cr & \simeq  -
\frac{\Lambda}{6} \sin \alpha_m\, r_L (r_S - r_L) ,}
\label{eq:final}
\]

\noindent where $r_L$ and $r_S$ are the distances to the lens and
source respectively. The angle between the
observer's line of sight and the path of the photon is denoted by
$\phi$, and we have used $r \sin \phi = r_L \sin \alpha_m$.

To clarify, $\alpha_\Lambda$ corresponds to the extra angle through which the photon appears to be deflected when tracking its motion in a physical coordinate system in the frame of $O$. The local deflection $\alpha_m$ occurs at the closest approach to $L$, while the extra $\Lambda$ contribution is a cumulative effect. The $\Lambda$-dependent terms given in (\ref{eq:alpha2}) and (\ref{eq:final}) appear particularly problematic since they do not vanish when the lensing mass is taken to be zero. Ordinarily the local deflection $\sin \alpha_m$  would vanish, yet it could equally have been a mirror we
used to deflect the photon, maintaining a non-zero $\alpha_m$ in an otherwise pure de Sitter
universe.

\begin{figure}
\includegraphics[width=80mm]{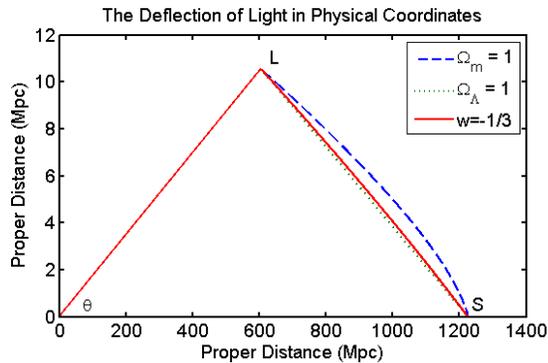} %coords.png
\caption{A comparison of a photon's path in different flat
cosmologies, using the physical distances from the perspective of
the observer at the origin. The lens $L$ and source $S$ appears at
the same angular-diameter distance in all cases, and the true
deflection angle is taken to be $\alpha=1^\circ$.
 The lines correspond to de Sitter ($\Omega_\Lambda=1$, dotted),
Einstein-de Sitter ($\Omega_m=1$, dot-dash), and the neutral
$w=-1/3$ cosmology  ($\Omega_m=2/3$, $\Omega_\Lambda=1/3$,
solid).} \label{fig:coords}
\end{figure}

How can we reconcile this modified deflection angle with the
concept that light travels in straight lines within FRW models?
The difference is highlighted in Figure \ref{fig:coords} where we
see the influence of our coordinate system. The lines plotted in
Figure \ref{fig:coords} have a physical interpretation as those which
would be marked on a sheet of graph paper centred on the observer.
Aside from the deflection, the photon follows a Euclidean
trajectory in comoving space, since the geometry is conformally
flat. However when we map this trajectory onto a coordinate system with proper distances -- that
measured by a ruler -- the transverse motion of a photon appears
bent by the acceleration of the cosmology, and the deflection
angle induced at $L$ appears to change.  Yet the observable angle
$\theta$ remains constant, as does the deflection angle for a set
of comoving observers, since we have fixed the angular diameter distances. Note that in de Sitter space, the same
distance corresponds to a substantially lower redshift, so the
magnitude of the apparent bending is reduced.

Now we quantify the relation between the angle in a comoving
coordinate system, and that observed in terms of proper distances.
Consider a photon at a comoving coordinate $(x,y)$ travelling at a
small angle $\alpha$ with respect to the radial direction (which
we take to be the $x$-axis, so $y=0$). The proper distance of the
photon, and its derivative with respect to the scale factor, is
given by

\[
\eqalign{ x_p &= a x \cr x_p' &= x + ax' }
\]

\noindent and similarly for $y$. The angle of the trajectory in
terms of comoving and proper distances is given by

\[
\eqalign{ \tan \alpha &= \frac{y'}{x'} , \cr \tan \alpha_p &=
\frac{y+ay'}{x + ax'} .}
\]

\noindent To first order in $x /ax'$ this leads to

\[
\eqalign{ \tan \alpha_p &= \frac{y'}{x'} - \frac{y'x}{ax'^2} \cr
&=  \left(1-\frac{x}{ax'} \right) \tan \alpha .}
\]

Note that for $\alpha = 0$ then $\alpha_p= 0$, and as we expect,
radial trajectories remain radial in terms of proper distances.
This may be simplified to leave

\[
\tan \alpha_p =  \left(1 - r\frac{da}{dt}\right) \tan \alpha ,
\]

\noindent For a homogeneous cosmology with equation of state $w$
then this may  be expressed as

\[
\tan \alpha_p =  (1 - rH_0 a^{-\frac{1+3w}{2}}) \tan \alpha ,
\]

\noindent highlighting the significance of $w=-1/3$,  which
corresponds to a cosmology with zero acceleration.

The photon trajectories plotted in Figure \ref{fig:coords} appear
distorted as a result of the aforementioned coordinate
transformation. However we stress that any observable such as
$\theta$ will remain unaltered, since only distant motion with
some transverse component may be affected. The deflection angle is
modified, as given by (\ref{eq:final}), yet remains unobservable
since we cannot measure the angle of emission. We consider the
physical interpretation of these results in the following section.

%The modified deflection angle from (\ref{eq:final}) is compensated
%by the angle of emission at $S$.

\sec{Aberration and the Origin of the Ishak--Rindler term}
\label{sec:origins}

\begin{figure}  \label{fig:box}
\includegraphics[width=80mm]{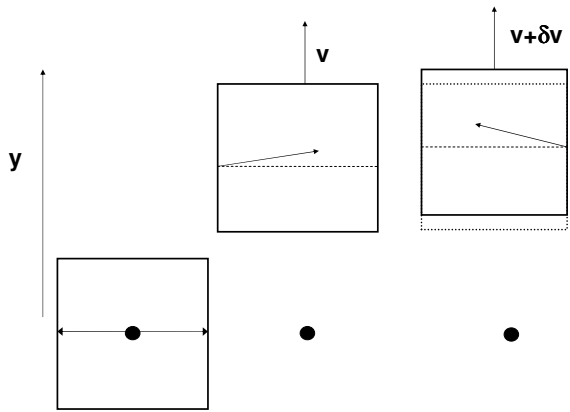} 
\caption{The bending of a photon trajectory in the absence of any lensing mass, in  de Sitter space. To begin on the left we sit in the frame of a reflective box and observe the photon bouncing back and forth. The central illustration shows the appearance of the same box now receding from some displaced viewpoint. As the box accelerates in de Sitter space, the angle of incidence must enlarge to match this, giving the illusion of lensing. This demonstrates the main problem associated with measuring the deflection of light in the rest frame of the lens.} 
\end{figure}

To illustrate the meaning of this bent trajectory, consider a
photon reflecting off the interior walls of a box
of proper size $\ell$ within a de
Sitter background. In the frame of the box the photon simply
bounces back and forth, only subject to a small degree of blue-
and red-shifting depending on which side of the box we sit.  Yet
for an observer located at a distance $y$ transverse to this
motion, as illustrated in Figure \ref{fig:box}, both the box and photon must appear to be accelerated. This
can be thought of in terms of an angular aberration, which to
first order in $v$ and $\theta$ is given by

\[ \label{eq:aberr}
\theta'  =  \theta(1+v_x) + v_y ,
\]

\noindent where $v_x$ and $v_y$ denotes the vertical and
horizontal velocities of the frame of $S'$ with respect to $S$,
and the photon is travelling in the positive $x$ direction.

The proper distance $y$ is simply related to the comoving
coordinate $\chi$ by

\[ \label{eq:comove}
y(t) = \chi  a(t)  .
\]

\noindent While the value of $\chi$ between two points remains
fixed, our distance $y$ changes at a rate governed  by the
Friedmann equations.

\[
\eqalign{ \ddot{y}  &= \chi \ddot{a} \cr &= \chi a  \left(-\frac{4
\pi \rho_m}{3} + \frac{\Lambda}{3} \right) . }\label{eq:comove2}
\]

For de Sitter space $\rho_m = 0$ and so $v_y = Hy$. Therefore at
the first bounce the photon appears to have a trajectory given by

\[ \label{eq:aberr2}
\theta'  =  \sqrt{\frac{\Lambda}{3}} \chi ,
\]

\noindent but accumulates an additional vertical
velocity after travelling a distance $\ell$ across the width of the box,

\[ \label{eq:comove3}
\eqalign{ v_y & = v_0 + \int \ddot{y}\, \ud t \cr &=
\sqrt{\frac{\Lambda}{3}} \chi + \frac{\Lambda \ell \chi}{3} ,}
\]

\noindent provided $v \ll c$ and $\ell \ll \chi$. The photon's angle of incidence
exceeds the angle of the previous bounce, and this is interpreted
as a bend angle of

\[ \label{eq:comove4}
\alpha_\Lambda  =  \frac{\Lambda \ell \chi}{3}.
\]

\noindent This deflection is actually an angular aberration arising from acceleration, due
to the change in velocity attained
by a freefalling body at a distance $\chi$. Note that the term
relates to that in (\ref{eq:alpha2}), but here the bend angle has
been defined as positive since in this case the photon is receding
from the observer. The analogy here is that one side of the box is the source, the other side the observer, and the reference frame is the lens (which in this case is massless). In the frame of the box - be it source or observer - no deflection is observed, yet the lens frame shows this anomalous bending.

Switching between different inertial frames therefore leaves us with
an angular aberration associated with the relative velocity
between the two frames. But of what physical significance are the
aberrations? They are ``real" angles in the sense that if a
physical tube were to be constructed down which light could be
shone over cosmological distances, it would need to be bent in
this manner. However it would also be in a non-inertial reference
frame, as the tube would need to be continuously accelerated in
order to counteract the gravitational forces which would otherwise
have led to it joining the Hubble flow. Therefore locally it would
appear that the photon is travelling straight while the bent
physical structure is accelerated in just the right manner so as to allow the photon to pass. Despite the temptation to consider the intuitive physical
coordinates, this example illustrates that it is much more natural
to think in terms of a comoving framework, such that a fixed point
corresponds to an appropriate inertial reference frame.

Recently Sereno \cite{2008arXiv0807.5123S} highlighted the
presence of an additional term given by $2mb\Lambda/3$. In
contrast to the aberration terms outlined above, which appear to
reduce the deflection angle due to cosmic acceleration, this
increases the deflection angle due to the cosmic expansion rate.
In the Appendix we present a heuristic approach that provides a
physical interpretation of the Sereno term. Essentially any
transverse motion by the lens will modify the deflection angle,
due to the time-dependent impact parameter. This term is found to
be consistent with the motion associated with the cosmological
expansion. One might worry that peculiar velocities may modify the
cosmic shear signal via this mechanism, though this contribution
has been shown to be too small to be of concern
\cite{2008arXiv0810.0180B}.

\sec{Angular-Diameter Distance} \label{sec:worked}

\begin{figure}[t]
\includegraphics[width=80mm]{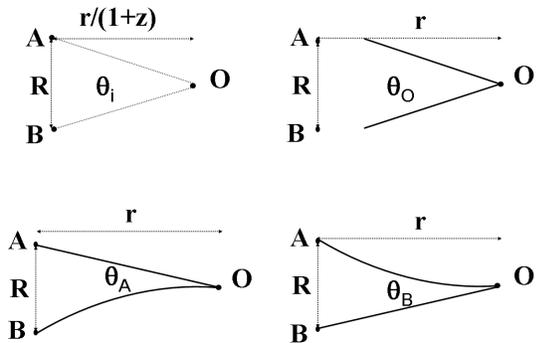}
\caption{Photons from sources $A$ and $B$, located at either end
of a rigid body of length $R$, propagate to the observer $O$ at a
proper distance $r$ in de Sitter space. As viewed from the
reference frame of $O$ (top-right), A (bottom-left), and B (bottom
right). The top-left panel illustrates the configuration at the
time of emission. Note that from the perspective of the observer,
the distance traversed by the photons does not match the
source-observer separation. $r$.} \label{fig:worked}
\end{figure}

We have already established the angle at which the photon appears
to impact an observer, from a distant perspective. Now we assess
the relative appearance of a more physically meaningful angle, the
path crossing of two photons. The setup involves two sources $A$
and $B$ separated by a fixed distance $R$, and an observer $O$ at
a distance $r/(1+z) \gg R$. Therefore the initial angle of
interest is $\theta_i = R(1+z)/r$, as illustrated in Figure
\ref{fig:worked}. By the time the photon reaches $O$, the physical
angle separating the bodies is $\theta_f = R/r$.

Consider the angle of incidence as determined by the source at
$B$ (bottom-right panel). In the time taken for the photon to travel the distance $r$,
the photon from $A$ acquires a vertical velocity, leading to a
bending angle such that

\[ \label{eq:std}
\theta_B  =  \theta_f - \frac{\Lambda R r}{3},
\]

\noindent From the perspective of $A$, its own photon now appears
straight yet the photon from $B$ is deflected, and by symmetry we
have $\theta_A = \theta_B$.

The relative motion of the comoving observer is

\[
v_x= \sqrt{\frac{\Lambda }{3}} r ,
\]

\noindent so once again by utilising (\ref{eq:aberr}) we arrive at

\[
\eqalign{ \theta_O & =  \theta_f (1 + v_x) \cr & =  \theta_f
\left(1 + \sqrt{\frac{\Lambda }{3}} r \right) \cr & =  R (1+z) /r ,
}
\]

\noindent where the redshift $z$ represents the horizontal recession
velocity. So we recover the expression for the angular-diameter
distance.

Alternatively, consider an observer on $A$ studying the angle at
which a telescope on $O$ is pointing in order to detect the source
at $B$. What angle must the telescope be pointed in order to let
the photon pass? The Lorentz contraction of the telescope actually
increases its apparent inclination, though this is an $O(v^2)$
effect. The problem is resolved simply by the motion of the
telescope, which allows the photon to pass at an angle of
approximately $\theta_t=\theta_p (1+v)$. This process provides a
alternative picture of how the factor of $(1+z)$ arises in the
angular-diameter distance.

\sec{Strong Lambda} \label{sec:strong}

As an aside, since we have been considering the limit of a weak
 field, let us address the stronger regime. If $\Lambda$ were to modify the deflection of
light, this would be expected to become most apparent in a
scenario were the length scales involved exceeded the event
horizon,  $r_\Lambda = \sqrt{3/\Lambda}$. Naturally the
Source-Observer distance is restricted to a sub-horizon scale, but
suppose a mass was positioned with a large transverse displacement
$R$, beyond the event horizon. One might be tempted to believe
that, for reasons of causality, the event horizon ``shields" the
photon from the distant mass, thereby nulling the lens. This is
evidently not the case, as the Kottler metric is still valid for
the regime $r>\sqrt{3/\Lambda}$, in much the same way as a black
hole gravitates beyond its event horizon. Causality is preserved
since no information is transmitted -- for instance any
gravitational waves emitted by the mass will remain confined
within the horizon. Reassuringly, this scenario also suggests that
in the context of gravitational lensing, no terms involving
$\Lambda r^2$ arise.

\sec{Conclusions} \label{sec:conc}

In this work we have placed light bending by a spherically
symmetric mass distribution with Lambda into a cosmological
context, and attempted to reconcile the apparent bending of light
as described in
\cite{2007PhRvD..76d3006R,2008PhRvD..77d3004S,2007arXiv0711.0673L,2008A&A...484..103S,2008GReGr.tmp...86S,2008arXiv0801.3514I,2007arXiv0710.4726I,2008arXiv0805.1630S,2008arXiv0807.0380S,2008arXiv0807.5123S},
with the conventional view that the cosmological constant does not
directly influence gravitational lensing.

To confirm that the cosmological constant does not contribute at
linear order to the deflection of light by a density fluctuation,
we explicitly transformed a perturbed FRW metric into the Kottler
metric. In the former metric, the linear lensing potential has no
explicit dependence on $\Lambda$, so the $\Lambda$-dependent
bending claimed to exist in the Kottler metric appears to be a
gauge artefact, with no direct implications for observations.

The source of explicit $\Lambda$-dependence primarily arises by
adopting physical distances to define the angles, and doing so in
the frame of reference of the lens rather than the observer merely
exacerbates the problem. Whilst physical scales provide an
intuitive picture of the photon's trajectory, it fails to take
into account the relative motion between local and distant
comoving observers, and the frame-dependence of the metric.

Terms involving $\Lambda r^2$ essentially arise from measuring the
photon's trajectory within a non-inertial frame of reference --
that of a particular physical coordinate, within the context of an
accelerating cosmology. This effect should therefore be considered
distinct from genuine gravitational lensing effects, where the
deflection angle is gauge invariant.

Broadly, we are in agreement with the analyses of Sereno \cite
{2008PhRvD..77d3004S,2008arXiv0807.5123S} in concluding that at
linear order, there is no influence of Lambda on light bending.
Indeed, for light bending on cluster scales, Sereno \cite
{2008arXiv0807.5123S} shows that the influence of Lambda on the
bend angle is third-order in the two mass- and Lambda-related
small quantities he introduces (of comparable size for clusters).
This term is several orders of magnitude smaller than second-order
mass terms which are routinely neglected.

Of course, the cosmological constant does still influence the lens
geometry, and it is primarily this modification to the
distance-redshift relation which allows weak lensing surveys to
constrain dark energy models. Our belief is that this application
can proceed without requiring modification of the basic lensing
theory.

\noindent{\bf Acknowledgements}\\
FS was funded by an STFC rolling grant, and is grateful for the
generous welfare support from the University of Edinburgh.  We
particularly thank Wolfgang Rindler and Mustapha Ishak for many
stimulating discussions.

\appendix

\sec{Lenses in Motion} \label{sec:app}

In the case of Newtonian Gravitational Lensing, we can evaluate
the deflection angle of a particle of arbitrary velocity by
considering the integrated force perpendicular to the direction of
motion.

Consider a particle of horizontal motion $v_x=v$ and initial
vertical motion $v_y=0$. After passing a gravitating mass the
particle accrues a vertical velocity determined by

\[
v_y = \int F_y \ud t  ,
\]

\noindent where

\[
F_y =  \frac{GM}{r^2} \sin{\theta}  .
\]

A change of variable leads to

\[ \eqalign{
v_y = &  \frac{GM}{v} \int_{-\infty}^\infty \frac{R}{r^3} \ud x  .
\label{eq:newt} \cr = & \frac{2GM}{Rv}}
\]

\noindent since $r^2 = R^2 + x^2$. For small angles, the Newtonian
deflection is then simply given by

\[ \eqalign{
\alpha_N = &  \frac{v_y}{v_x} \cr = & \frac{2GM}{Rv^2}}
\]

For the case $v=c$, this result is of course half that predicted
by General Relativity, and this discrepancy can be attributed to
the matching contribution from the spatial component of the
metric, which is otherwise negligible for $v \ll c$. The Newtonian
approach allows us to gain some intuition on the influence of a
lens in transverse motion. A lens travelling with a velocity $v_L$
introduces a time-dependence to the vertical displacement $R$
which we parameterise as $R=R_0 + v_L x$ and we have set $v=1$. In
this case we find (\ref{eq:newt}) becomes

\[ \eqalign{
v_y = &  GM \int_{-\infty}^\infty \frac{R_0 + v_L x}{\left[ (R_0 +
v_L x)^2 + x^2\right]^{3/2}} \ud x  , \cr = & \frac{2GM}{R
\sqrt{1-v_L^2}}}  .
\]

This corresponds to a modification to the Newtonian deflection
angle given by

\[
\delta \alpha_N \simeq   \frac{GMv_L^2}{R}
\]

If we define the velocity to correspond to the Hubble flow, then
the transverse lens velocity is given by $v_L = HR$,  and
reintroducing the factor of two from General Relativity leaves us
with

\[
\delta \alpha \simeq   2GMRH^2 .
\]

\noindent Finally taking $H^2 = \Lambda/3$ corresponds to the term
from Sereno

\[
\delta \alpha \simeq   2GMR \Lambda/3 .
\]

In most practical cases the peculiar motion of the lens will
likely far exceed this influence.
\bibliography{M:/Routines/dis}
\end{document}